\documentclass[useAMS,usenatbib]{mn2e}
\usepackage{txfonts}
\usepackage{natbib}
\usepackage[all]{xy}
\usepackage[dvips]{graphicx}

\def\del#1{{}}

\newcommand{\ltsima}{$\; \buildrel < \over \sim \;$}
\newcommand{\lsim}{\lower.5ex\hbox{\ltsima}}
\newcommand{\gtsima}{$\; \buildrel > \over \sim \;$}
\newcommand{\gsim}{\lower.5ex\hbox{\gtsima}}
\newcommand{\bra}{\langle}
\newcommand{\ket}{\rangle}

\newcommand{\dd}{\mathrm{d}}

\newcommand{\veck}{\bmath{k}}

\title[iSW-effect and bias evolution]
{Implications of bias evolution on measurements of the integrated Sachs-Wolfe effect: errors and biases in parameter estimation}
\author[B.M. Sch{\"a}fer, M. Douspis, N. Aghanim]
{Bj{\"o}rn Malte Sch\"afer$^{1,2}$\thanks{e-mail: bjoern.malte.schaefer@ita.uni-heidelberg.de}, Marian Douspis$^1$, Nabila Aghanim$^1$\\
$^1$Institut d'Astrophysique Spatiale, Universit{\'e} de Paris XI, b{\^a}timent 120-121, Centre universitaire d'Orsay, 91400 Orsay CEDEX, France\\
$^2$Astronomisches Recheninstitut, Zentrum f{\"u}r Astronomie, Universit{\"a}t Heidelberg, M{\"o}nchhofstra{\ss}e 12, 69120 Heidelberg, Germany}

\begin{document}
\pagerange{\pageref{firstpage}--\pageref{lastpage}}
\pubyear{2008}
\maketitle
\label{firstpage}

% --- abstract --- %
\begin{abstract}
The subject of this paper is a quantification of the impact of uncertainties in bias and bias evolution on the interpretation of measurements of the integrated Sachs-Wolfe effect, in particular on the estimation of cosmological parameters. We carry out a Fisher-matrix analysis for quantifying the degeneracies between the parameters of a dark energy cosmology and bias evolution, for the combination of the PLANCK microwave sky survey with the EUCLID main galaxy sample, where bias evolution $b(a)=b_0+(1-a)b_a$ is modelled with two parameters $b_0$ and $b_a$. Using a realistic bias model introduces a characteristic suppression of the iSW-spectrum on large angular scales, due to the altered distance-weighting functions. The errors in estimating cosmological parameters if the data with evolving bias is interpreted in the framework of cosmologies with constant bias is quantified in an extended Fisher-formalism. We find that the best-fit values of all parameters are shifted by an amount comparable to the statistical accuracy: The estimation bias in units of the statistical accuracy amounts to 1.19 for $\Omega_m$, 0.27 for $\sigma_8$, and 0.72 for $w$ for bias evolution with $b_a=1$. Leaving $b_a$ open as a free parameter deteriorates the statistical accuracy, in particular on $\Omega_m$ and $w$.
\end{abstract}

% --- keywords --- %

% --- section: introduction --- %
\section{Introduction}
The integrated Sachs-Wolfe (iSW) effect \citep{1967ApJ...147...73S, rees_sciama_orig, 1994PhRvD..50..627H, 2002PhRvD..65h3518C, 2006MNRAS.369..425S}, which refers to the frequency change of cosmic microwave background (CMB) photons if they cross time evolving gravitational potentials, is a direct probe of dark energy because it vanishes in cosmologies with $\Omega_m=1$ \citep{1996PhRvL..76..575C}. By now, it has been detected with high significance with a number of different tracer objects \citep{2003ApJ...597L..89F, 2004Natur.427...45B, 2004ApJ...608...10N, 2005PhRvD..72d3525P, 2006PhRvD..74f3520G, 2006PhRvD..74d3524P, 2006MNRAS.365..171G, 2006MNRAS.372L..23C, 2006MNRAS.365..891V, 2007MNRAS.377.1085R, 2007MNRAS.376.1211M, 2008arXiv0801.4380G}.

Up to now, the aforementioned studies assumed a constant bias in the interpretation of the signal, although it is well established by theoretical studies that the bias of the tracer objects evolves by as much as 50\% from redshift unity to today. Bias and in particular bias evolution has been the topic of a number of papers, both analytical and numerical. \citet{fry96}, \citet{tegmark98} and \citet{2001ApJ...550..522B} looked at a two component fluid composed of dark matter and galaxies, and derived models for bias evolution in perturbation theory: In their calculations the bias undergoes a slow evaluation, which is linked to the linear growth function, and decreases slowly towards unity. Numerical investigations have found equivalent results \citep[e.g.][]{blanton00}. \citet{2008arXiv0801.0642H} apply iSW-tomography to measurements of the cross-spectrum $C_{\tau\gamma}(\ell)$ and the tracer autocorrelation $C_{\gamma\gamma}(\ell)$ using families of tracer objects peaking at different redshifts for controlling uncertainties in the tracer redshift distribution, aiming at disinguishing flat and curved cosmologies with the iSW-effect. Revisiting the iSW-detection in the NVSS radio source catalogue, \citet{2008MNRAS.386.2161R} point out effects of bias evolution on the interpreation of the iSW-signal. In their case the observational bias (due to luminosity evolution) can correct the discrepancies for the iSW-prediction for a $\Lambda$CDM model which arose by using an updated redshift distribution for the tracer objects.

We set out to perform a Fisher-analysis to quantify the magnitude of varying tracer bias on iSW-measurements and to investigate the basic degeneracies between the cosmological model, and in particular the dark energy equation of state properties, and cosmological bias evolution. We provide a quantification of degeneracies for the combination of the PLANCK CMB-observation with the EUCLID galaxy sample. In a further step, we investigate the error in parameter estimation if the data with evolving bias is interpreted in terms of a cosmology with constant bias, using the extension of the Fisher-matrix formalism worked out by \citep{2009MNRAS.392.1153T}, with focus on the estimation of the dark energy equation of state parameter $w$.

After compiling the key formul{\ae} describing structure formation in dark energy cosmologies in Sect.~\ref{sect_homogeneous}, we introduce the two relevant observational channels in Sect.~\ref{sect_channels} and derive the spectrum $C_{\tau\gamma}(\ell)$ between iSW-temperature perturbation $\tau$ and the galaxy density $\gamma$ in Sect.~\ref{sect_isw}. We quantify the degeneracies between the cosmological parameters using a Fisher-matrix analysis in Sect.~\ref{isw_fisher} and extend this formalism to describe the parameter estimation bias resulting from interpreting the iSW-signal with bias evolution in a model with constant bias in Sect.~\ref{isw_bias}. A summary of our results is given in Sect.~\ref{sect_summary}. 

As cosmologies, we consider the family of spatially flat homogeneous dark energy models with constant dark energy equation of state, and with Gaussian adiabatic initial conditions in the cold dark matter field. Specific parameter choices for the fiducial model in the Fisher-matrix analysis are $H_0=100h \:\mathrm{km}/s/\mathrm{Mpc}$ with $h=0.72$, $\Omega_m=0.25$, $\sigma_8=0.8$, $w=-1.0$ and $n_s=1$, with a non-evolving bias parameter of unity.

% --- section: key formulae --- %
\section{Cosmology and structure formation}\label{sect_homogeneous}

% --- subsection: dark energy cosmologies --- %
\subsection{Dark energy cosmologies}
In a spatially flat dark energy cosmology with the matter density parameter $\Omega_m$, the Hubble function $H(a)=\dd\ln a/\dd t$ is given by
\begin{equation}
\frac{H^2(a)}{H_0^2} = \frac{\Omega_m}{a^{3}} + (1-\Omega_m)\exp\left(3\int_a^1\dd a\:\frac{1+w(a)}{a}\right),
\end{equation}
with the dark energy equation of state $w(a)$. The value $w\equiv -1$ corresponds to the cosmological constant $\Lambda$. The relation between comoving distance $\chi$ (given in terms of the Hubble distance $d_H=c/H_0$) and scale factor $a$ is given by
\begin{equation}
\chi = c\int_a^1\dd a\:\frac{1}{a^2 H(a)},
\end{equation}
with the speed of light $c$. The dark energy equation of state $w(a)$ can be parameterised by its first order Taylor expansion with respect to the scale-factor $a$ \citep{2001IJMPD..10..213C, 2003MNRAS.346..573L},
\begin{equation}
w(a) = w_0 + (1-a) w_a.
\end{equation}
The conformal time, which is related to the cosmic time $t$ by the differential $\dd\eta=\dd t/a$, follows directly from the definition of the Hubble function,
\begin{equation}
\eta = \int_a^1\dd a\: \frac{1}{a^2H(a)},
\end{equation}
in units of the Hubble time $t_H=1/H_0$. Hence, $\eta$ is defined in complete analogy to the comoving distance $\chi$.

% --- subsection: CDM spectrum --- %
\subsection{CDM power spectrum}
Inflationary models suggest that the CDM power spectrum $P(k)$, which describes the fluctuation statistics of the Gaussian density field, $\bra\delta(\veck)\delta(\veck^\prime)\ket=(2\pi)^3\delta_D(\veck+\veck^\prime)P(k)$ in the case of homogeneous and isotropic fluctuations, can be written
\begin{equation}
P(k)\propto k^{n_s} T^2(k),
\end{equation} 
with the transfer function \citep{1986ApJ...304...15B},
\begin{displaymath}
T(q) = \frac{\ln(1+2.34q)}{2.34q}\left(1+3.89q+(16.1q)^2+(5.46q)^3+(6.71q)^4\right)^{-\frac{1}{4}},
\end{displaymath}
where the wave vector $q$ is given in units of the shape parameter $\Gamma\simeq\Omega_m h$. $P(k)$ is normalised to the value $\sigma_8$ on the scale $R=8~\mathrm{Mpc}/h$,
\begin{equation}
\sigma_R^2 = \frac{1}{2\pi^2}\int\dd k\: k^2 W^2(kR) P(k),
\end{equation}
with a Fourier-transformed spherical top-hat $W(x)=3j_1(x)/x$ as the filter function. $j_\ell(x)$ denotes the spherical Bessel function of the first kind of order $\ell$ \citep{1972hmf..book.....A}.

% --- subsection: structure growth --- %
\subsection{Structure growth in dark energy cosmologies}
The homogeneous growth of the overdensity field, $\delta(\bmath{x},a)=D_+(a)\delta(\bmath{x},1)$ is described by the growth function $D_+(a)$, which is a solution to the differential equation \citep{1998ApJ...508..483W, 1997PhRvD..56.4439T, 2003MNRAS.346..573L},
\begin{equation}
\frac{\dd^2}{\dd a^2}D_+
+\frac{1}{a}\left(3+\frac{\dd\ln H}{\dd\ln a}\right)\frac{\dd}{\dd a}D_+ = 
\frac{3}{2a^2}\Omega_m(a)D_+(a).
\label{eqn_growth}
\end{equation}
In the standard cold dark matter (SCDM) cosmology with $\Omega_m=1$ and $3+\dd\ln H/\dd\ln a=\frac{3}{2}$, the solution for $D_+(a)$ is simply the scale factor itself, $D_+(a)=a$. This solution determines the initial conditions $D_+(0)=0$ and $\dd/\dd a D_+(0)=1$, due to early-time matter domination and causes the iSW-effect to vanish in $\Omega_m=1$ cosmologies.

% --- subsection: bias and bias evolution --- %
\subsection{Bias and bias evolution}
Dark matter haloes hosting galaxies form by spherical collapse at peaks in the cosmic density field. The fractional perturbation $\Delta n/\bra n\ket$ in the spatial number density $n$ of galaxies is related to the overdensity $\delta=\Delta\rho/\rho$ by the bias parameter \citep{1986ApJ...304...15B},
\begin{equation}
\frac{\Delta n}{\bra n\ket} = b\frac{\Delta\rho}{\rho}.
\end{equation}
The bias parameter $b$ is in general scale dependent and subjected to a time-evolution, as it slowly decreases with time towards unity. \citet{fry96} and \citet{tegmark98} consider the cosmologiecal evolution of bias in perturbation theory for small perturbations $\Delta n/n\ll 1$ and find
\begin{equation} 
b_\mathrm{lin}(a) = \frac{D_+(a)-1+b_0}{D_+(a)},
\label{eq:b_evol}
\end{equation}
with the growth function $D_+(a)$ and the value $b_0$ of the bias parameter today. Motivated by their findings, we use a simple, linear model for bias evolution with two parameters $b_0$ and $b_a$,
\begin{equation}
b(a) = b_0 + (1-a) b_a,
\end{equation}
in order to accomodate effects from bias evolution if the galaxy density is strongly perturbed \citep{blanton00, 2001ApJ...550..522B} and from velocity bias \citep{2008MNRAS.385L..78P}. In the course of structure formation, the bias of any object tends to unity, $b_0=1$, as first shown by \citet{fry96}, which we will assume applies to the galaxy sample. By comparing the observed galaxy correlation function in the VMOS-VLT survey with the dark matter correlation function derived from $n$-body simulations, \citet{2005A&A...442..801M} propose the approximation to the bias evolution law,
\begin{equation}
b(z) = 1 + (0.03\pm0.01)(1+z)^{3.3\pm0.6},
\end{equation}
which if linearised for values of $a$ in the vicinity of 1, yields as a lower limit a value of $b_a\simeq0.1$.

% --- section: observational channels --- %
\section{Line of sight expressions}\label{sect_channels}

% --- subsection: iSW-effect in coupled cosmologies  --- %
\subsection{iSW-effect and large-scale structure tracers}
The iSW-effect is caused by gravitational interaction of a CMB photon with a time-evolving potential $\Phi$. The fractional perturbation $\tau$ of the CMB temperature $T_\mathrm{CMB}$ is given by \citep{1967ApJ...147...73S, rees_sciama_orig}
\begin{equation}
\tau 
= \frac{\Delta T}{T_\mathrm{CMB}} 
\equiv \frac{2}{c^2}\int\dd\eta\: \frac{\partial\Phi}{\partial\eta} 
= -\frac{2}{c^3}\int_0^{\chi_H}\dd\chi\: a^2 H(a) \frac{\partial\Phi}{\partial a},
\label{eqn_sachs_wolfe}
\end{equation}
where $\eta$ denotes the conformal time. In the last step, we have replaced the integration variable by the comoving distance $\chi$, which is related to the conformal time by $\dd\chi = -c\dd\eta = -c\dd t/a$, and the time derivative of the growth function has been rewritten in terms of the scale factor $a$, using the definition of the Hubble function $\dd a/\dd t = aH(a)$, with the cosmic time $t$. The gravitational potential $\Phi$ follows from the Poisson equation in the comoving frame, 
\begin{equation}
\Delta\Phi = 4\pi G a^2\rho_m(a)\:\delta
\end{equation}
where Newton's constant $G$ is replaced with the critical density $\rho_\mathrm{crit}(a)=3H^2(a)/(8\pi G)$, $\rho_m(a) = \Omega_m(a)\rho_\mathrm{crit}(a)$,
\begin{equation}
\Delta\Phi = \frac{3H_0^2\Omega_m}{2a}\delta.
\end{equation}
Substitution yields a line of sight expression for the linear iSW-effect $\tau$ (integrating along a straight line and using the flat-sky approximation), sourced by the linear density field $\delta$,
\begin{equation}
\tau = 
\frac{3\Omega_m}{c}\int_0^{\chi_H}\dd\chi\: 
a^2 H(a)\:\frac{\dd}{\dd a}\left(\frac{D_+}{a}\right)\:\frac{\Delta^{-1}}{d_H^2}\delta,
\end{equation}
with the inverse (dimensionless) Laplace operator $\Delta^{-1}/d_H^2$ solving for the (dimensionless) potential 
$\varphi$,
\begin{equation}
\varphi\equiv\frac{\Delta^{-1}}{d_H^2}\delta,
\end{equation}
with the Hubble distance $d_H=c/H_0$. Due to the achromaticity of the iSW-effect, a measurement of the temperature perturbations of the microwave sky can not distinguish between primary and secondary, iSW-induced anisotropies. For that reason, the iSW-effect is measured in cross-correlation with a tracer of the large-scale structure.

% --- subsection: galaxy density --- %
\subsection{Galaxy density}
The bias prescription $b(\chi)$ relates the galaxy number density to the density field $\delta$,
\begin{equation}
\gamma = \int_0^{\chi_H}\dd\chi\:p(z)\frac{\dd z}{\dd\chi} b(\chi) D_+(\chi)\:\delta.
\end{equation}
$p(z)\dd z$ is the redshift distribution of the surveyed galaxy sample, transformed in terms of the comoving distance $\chi$, and $D_+(\chi)$ is the growth function of the density field at the scale factor $a(\chi)$. We use the redshift distribution of the main galaxy sample of EUCLID \citep{2008arXiv0802.2522R}, which will observe half of the sky out to redshifts of order unity and which will be of particular use for iSW-observations \citep{2008arXiv0802.0983D}. In the parameterisation proposed by \citet{1995MNRAS.277....1S}, the redshift distribution is approximated by
\begin{equation}
p(z)\dd z = p_0\left(\frac{z}{z_0}\right)^2\exp\left(-\left(\frac{z}{z_0}\right)^\beta\right)\dd z
\quad\mathrm{with}\quad \frac{1}{p_0}=\frac{z_0}{\beta}\Gamma\left(\frac{3}{\beta}\right),
\end{equation}
with $\beta=3/2$ and $z_0=0.64$, which results in a median redshift of $z_\mathrm{med}=0.9$. We define the mean bias $\bar{b}$ in a bias evolution model $b(z)$,
\begin{equation}
\bar{b} = \int\dd z\: p(z) b(z),
\end{equation}
for a galaxy sample with the redshift distribution $p(z)\dd z$.

\begin{figure}
\resizebox{\hsize}{!}{\includegraphics{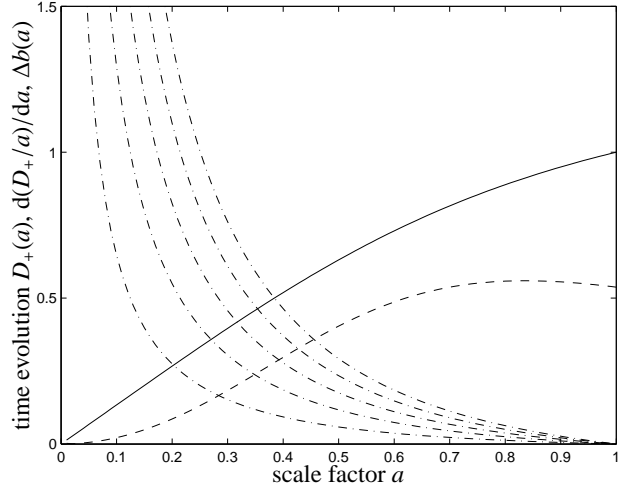}}
\caption{Time evolution of the source terms: $D_+(a)$ for the density field (solid line), $\dd (D_+/a)/\dd a$ of the iSW-effect (dashed line) and $\Delta b(a) = b_\mathrm{lin}(a)-b_0$ of the linear biasing model (dash-dotted, from top to bottom for $b_0=1.5,1.4,1.3,1.2,1.1$), with $\Lambda$CDM as the cosmological model.}
\label{figure_time_evolution}
\end{figure}

Fig.~\ref{figure_time_evolution} shows the time evolution of the source fields, i.e. the growth function $D_+(a)$ for the density field, and the time derivative of $D_+(a)/a$ for the gravitational potential. Additionally, the linearised bias evolution model $\Delta b(a) = b_\mathrm{lin}(a)-b_0$ is plotted, for a range of bias parameters $b_0$. As pointed out by a number of authors, the bias decreases with time and tends towards unity, although in the linear bias model (for small galaxy overdensities), the bias can never reach a value of exactly unity today if the galaxies form an initially biased population.

% --- section: iSW angular power spectra --- %
\section{Angular power spectra}\label{sect_isw}
In summary, the line of sight integrals for the iSW-temperature perturbation $\tau$ and the galaxy density $\gamma$ read:
\begin{eqnarray}
\tau & = & \frac{3\Omega_m}{c}\int_0^{\chi_H}\dd\chi\: a^2H(a)\frac{\dd}{\dd a}\left(\frac{D_+}{a}\right)\:\varphi,\\
\gamma & = & \int_0^{\chi_H}\dd\chi\: p(z)\frac{\dd z}{\dd\chi}b(\chi) D_+(\chi)\:\delta,
\end{eqnarray}
where we have defined the dimensionless potential $\varphi\equiv\Delta^{-1}\delta/d_H^2$, rescaled with the square of the Hubble distance $d_H=c/H_0$ for convenience. The weighting functions
\begin{eqnarray}
W_\tau(\chi) & = & \frac{3\Omega_m}{c}a^2 H(a) \frac{\dd}{\dd a}\frac{D_+}{a},\\
W_\gamma(\chi) & = & p(z)\frac{\dd z}{\dd\chi} b(\chi) D_+(\chi),
\end{eqnarray}
can be identified, which allow the expressions for the angular cross spectra to be written in a compact notation, applying a Limber-projection \citep{1954ApJ...119..655L} in the flat-sky approximation, for simplicity:
\begin{equation}
C_{\tau\gamma}(\ell) = \int_0^{\chi_H}\dd\chi\: \frac{W_\tau(\chi)W_\gamma(\chi)}{\chi^2}P_{\delta\varphi}(k=\ell/\chi)
\end{equation}
with the cross-spectrum $P_{\delta\varphi}(k) = P_{\delta\delta}(k) / (d_H k)^2$. The angular auto-spectra are given by:
\begin{eqnarray}
C_{\tau\tau}(\ell) & = & \int_0^{\chi_H}\dd\chi\: \frac{W_\tau^2(\chi)}{\chi^2}P_{\varphi\varphi}(k=\ell/\chi),\\
C_{\gamma\gamma}(\ell) & = & \int_0^{\chi_H}\dd\chi\: \frac{W_\gamma^2(\chi)}{\chi^2}P_{\delta\delta}(k=\ell/\chi),
\end{eqnarray}
which will be needed in the expression for the covariance of $C_{\tau\gamma}(\ell)$. In analogy to $P_{\delta\varphi}(k)$, the spectrum of the potential $\varphi$ is defined as $P_{\varphi\varphi}(k) = P_{\delta\delta}(k) / (d_H k)^4$.

\begin{figure}
\resizebox{\hsize}{!}{\includegraphics{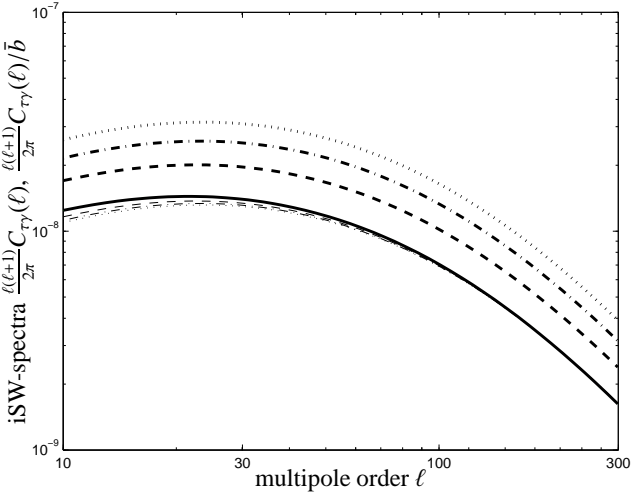}}
\caption{Angular iSW-spectra $C_{\tau\gamma}(\ell)$ (thick lines) and $C_{\tau\gamma}(\ell)/\bar{b}$ (thin lines) for the EUCLID main galaxy sample, for bias evolution parameters $b_a = 0$ (corresponding to the average bias $\bar{b}=1$, solid line), $b_a = 1$ ($\bar{b}= 1.46$, dashed line), $b_a = 2$ ($\bar{b}=1.93$, dash-dotted line) and $b_a = 3$ ($\bar{b}=2.39$, dotted line), for the EUCLID main galaxy sample with unit bias $b_0=1$ today.}
\label{fig_spectra}
\end{figure}

Angular cross power spectra $C_{\tau\gamma}(\ell)$ for a range of bias evolution paramters $b_a$ is depicted in Fig.~\ref{fig_spectra}. Assuming that an estimate of the average bias $\bar{b}$ follows from the spectrum $C_{\gamma\gamma}(\ell)$ via $\bar{b}\simeq C_{\gamma\gamma}/C_{\tau\gamma}$ or from other observational channels, the spectrum is rescaled with $\bar{b}$ because to first order, the iSW-spectrum $C_{\tau\gamma}(\ell)$ is proportional to the mean bias. Fig.~\ref{fig_spectra} suggests that on large angular scales, the difference in amplitude amounts to $7\ldots12\%$ for the range of $b_a$ values considered here, relative to the constant bias model. This difference decreases towards smaller angular scales, which can be easily understood by decomposing $W_\gamma(\chi)$ according to
\begin{equation}
W_\tau(\chi)W_\gamma(\chi) \equiv W^\prime(\chi) b(\chi) = W^\prime(\chi) \left[b_0 + (1-a)b_a\right].
\end{equation}
The iSW-spectrum then consists of two contributions,
\begin{equation}
C_{\tau\gamma}(\ell) = 
b_0\int_0^{\chi_H}\dd\chi\:\frac{W^\prime(\chi)}{\chi^2}P_{\delta\varphi} + 
b_a\int_0^{\chi_H}\dd\chi\:\frac{W^\prime(\chi)}{\chi^2}(1-a)P_{\delta\varphi},
\end{equation}
in analogy to the mean galaxy bias $\bar{b}$, 
\begin{equation}
\bar{b} = \int\dd z\: p(z) b(z) = b_0 + b_a \int\dd z\:p(z) (1-a),
\end{equation}
which is scale-independent and suppresses the amplitude in the scaled spectrum $C_{\tau\gamma}(\ell)/\bar{b}$, the suppression being proportional to $b_a/b_0$. The scaled angular spectrum, however, has in addition a scale-dependent contribution of the form
\begin{equation}
\frac{b_a}{b_0}\int_0^{\chi_H}\dd\chi\:\frac{W^\prime(\chi)}{\chi^2} (1-a) P_{\delta\varphi}(k=\ell/\chi),
\end{equation}
whose integrand is large at early times, i.e. small $a$, which corresponds to large comoving distances $\chi$. Consequently, CDM spectrum amplitudes corresponding to larger wave vectors $k$ get projected out, which leads to an overall decrease in the suppression, because $P_{\delta\varphi}(k)\propto P(k)/k^2$ is a monotonically decreasing function for the choice of the CDM transfer function $T(k)$. The unscaled spectra $C_{\tau\gamma}(\ell)$, on the contrary, vary by almost half an order of magnitude, indicating the relative importance of $b_0$ and $b_a$.

% --- section: Fisher-matrix analysis --- %
\section{Fisher-matrix analysis}\label{isw_fisher}

% --- subsection: Fisher-matrix for the iSW-measurements --- %
\subsection{Fisher-matrix for the iSW-spectrum $C_{\tau\gamma}(\ell)$}
The Fisher matrix, which quantifies the decrease in likelihood if a model parameter $x_\mu$ moves away from the fiducial value, can be computed for a local Gaussian approximation to likelihood $\mathcal{L}\propto\exp(-\chi^2/2)$. The Fisher-matrix for the measurement of $C_{\tau\gamma}(\ell)$ is given by
\begin{equation}
F_{\mu\nu}^\mathrm{iSW} = \sum_{\ell=\ell_\mathrm{min}}^{\ell_\mathrm{max}}
\frac{\partial C_{\tau\gamma}(\ell)}{\partial x_\mu}
\mathrm{Cov}^{-1}\left(C_{\tau\gamma}(\ell),C_{\tau\gamma}(\ell)\right)
\frac{\partial C_{\tau\gamma}(\ell)}{\partial x_\nu}.
\end{equation}
We construct the Fisher-matrix $F_{\mu\nu}^\mathrm{iSW}$ for $\Lambda$CDM as the fiducial cosmological model, where the parameter space is spanned by $\Omega_m$, $\sigma_8$, $h$, $n_s$, $w$ and $b_a$. The fiducial values for the parameters are $\Omega_m=0.25$, $\sigma_8=0.8$, $h=0.72$, $n_s=1$, $w=-1$ and $b_a=0$, i.e. a non-evolving bias. It should be emphasised at this point that in a measurement of $C_{\tau\gamma}(\ell)$ alone with $b_a=0$, the parameters $\sigma_8$ and $b_0$ are completely degenerate (leading to a singular Fisher-matrix),
\begin{equation}
\frac{\partial C_{\tau\gamma}(\ell)}{\partial\sigma_8} = 
\frac{2b_0}{\sigma_8}\frac{\partial C_{\tau\gamma}(\ell)}{\partial b_0},
\end{equation}
which is the reason why we set $b_0=1$, do not consider it as an independent parameter in the Fisher-analysis and absorb it into the normalisation of the spectrum $\sigma_8$. 

\begin{figure}
\resizebox{\hsize}{!}{\includegraphics{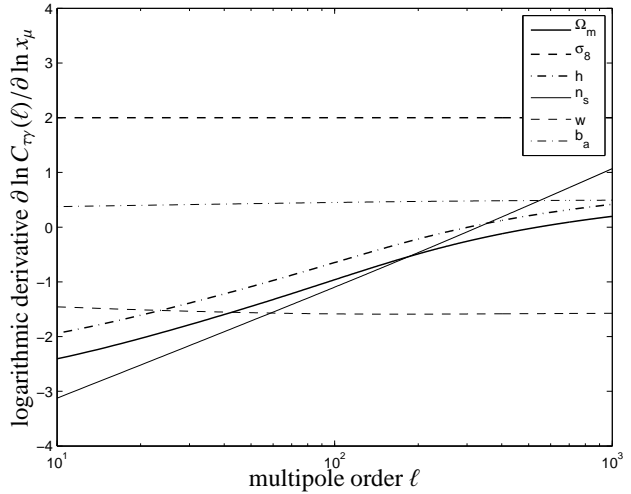}}
\caption{Logarithmic derivative $\partial\ln C_{\tau\gamma}(\ell)/\partial  x_\mu$ of the iSW-spectrum with respect to the cosmological parameters $x_\mu$, specifically $\Omega_m$ (thick solid line), $\sigma_8$ (thick dashed line), $h$ (thick dash-dotted line), $n_s$ (thin solid line), $w$ (thin dashed line) and $b_a$ (thin dash-dotted line).}
\label{fig_log_derivative}
\end{figure}

Implicitly, we assume priors on spatial flatness, $\Omega_m+\Omega_\Lambda=1$ and on the primoridal slope of the CDM spectrum $n_s=1$, and additionally neglect the weak dependence of the shape parameter on the baryon density $\Omega_b$. We extend the computation from $\ell_\mathrm{min}=10$ to $\ell_\mathrm{max}=200$, because at $\ell\lsim\ell_\mathrm{min}$ the small-angle approximation ceases to be applicable and because iSW-contributions from $\ell\gsim\ell_\mathrm{max}$ are small and would be dominated by the nonlinear iSW-effect. Fig.~\ref{fig_log_derivative} illustrates the sensitivity of the iSW-spectrum $C_{\tau\gamma}(\ell)$ on variations of the cosmological parameters, by plotting the logarithmic derivatives $\dd\ln C_{\tau\gamma}(\ell)/\dd\ln x_\mu$ (only in the case of the bias evolution parameter $b_a$ we give $\dd\ln C_{\tau\gamma}(\ell)/\dd b_a$ instead as the reference value $b_a=0$). One immediately recognises the degeneracy between $\sigma_8$ and $b_a$, which increases the mean bias, as well as between $\Omega_m$ and $h$ as they determine the CDM shape parameter.

% --- subsection: noise sources ---%
\subsection{Noise sources and covariance}
In an actual observation, the iSW-power spectrum is modified by the intrinsic CMB-fluctuations, the instrumental noise and the beam as a noise source, assuming mutual uncorrelatedness of the individual noise sources. The galaxy correlation function assumes a Poissonian noise term,
\begin{eqnarray}
\tilde{C}_{\tau\tau}(\ell) & = & C_{\tau\tau}(\ell) + C_\mathrm{CMB}(\ell) + w_T^{-1}B^{-2}(\ell),\\
\tilde{C}_{\gamma\gamma}(\ell) & = & C_{\gamma\gamma}(\ell) + \frac{1}{n},\label{eqn_obs_gg}
\end{eqnarray}
For PLANCK's noise levels the value $w_T^{-1}=(0.02\umu\mathrm{K})^2$ has been used, and the beam was assumed to be Gaussian, $B^{-2}(\ell)=\exp(\Delta\theta^2\:\ell(\ell+1))$, with a FWHM-width of $\Delta\theta = 7\farcm1$, corresponding to the $\nu=143$~GHz channel closest to the CMB-maximum. $n$ in eqn.~(\ref{eqn_obs_gg}) corresponds to the number of galaxies per steradian. EUCLID is expected to survey the entire extragalactic sky and to cover the solid angle $\Delta\Omega=2\pi$, i.e. $f_\mathrm{sky}=0.5$, yielding $4.7\times10^8$ galaxies per steradian. The observed cross power spectra are unbiased estimates of the actual spectra,
\begin{equation}
\tilde{C}_{\tau\gamma}(\ell) = C_{\tau\gamma}(\ell),
\end{equation}
in the case of uncorrelated noise terms. For the determination of the spectrum $C_\mathrm{CMB}(\ell)$ of the intrinsic CMB anisotropies, the CAMB code written by \citet{2000ApJ...538..473L} was employed. The covariance of the spectrum $C_{\tau\gamma}(\ell)$ is given in terms of the observed spectra $\tilde{C}_{XY}(\ell)$, with $X,Y\in \left\{\tau,\gamma\right\}$, which follow directly from applying the Wick-theorem,
\begin{equation}
\mathrm{Cov}(C_{\tau\gamma},C_{\tau\gamma}) = 
\frac{1}{2\ell+1}\frac{1}{f_\mathrm{sky}}
\left[\tilde{C}_{\tau\gamma}^2(\ell) + \tilde{C}_{\tau\tau}(\ell)\tilde{C}_{\gamma\gamma}(\ell)\right],
\end{equation}
with a cosmic variance error $\propto 1/f_\mathrm{sky}$. Due to the relatively small signal to noise ratio attainable by iSW-measurements (the largest contribution to the covariance being the primary CMB anisotropies) we enhance the measurement by CMB priors,
\begin{equation}
F_{\mu\nu} = F_{\mu\nu}^\mathrm{iSW} + F_{\mu\nu}^\mathrm{CMB},
\end{equation}
because $C_{\tau\gamma}(\ell)$ and $C_\mathrm{CMB}(\ell)$ constitute independent measurements. The Fisher matrix describing the CMB parameter bounds is marginalised over the optical depth $\tau$ and the baryon density $\Omega_b$.

% --- subsection: parameter bounds and degeneracies --- % 
\subsection{Parameter bounds and degeneracies}
The $\chi^2$-function for a pair of parameters $(x_\mu,x_\nu)$ can be computed from the inverse $(F^{-1})_{\mu\nu}$ of the Fisher matrix,
\begin{equation}
\chi^2 = 
\left(
\begin{array}{c}
\Delta x_\mu \\
\Delta x_\nu
\end{array}
\right)^t
\left(
\begin{array}{cc}
(F^{-1})_{\mu\mu} & (F^{-1})_{\mu\nu} \\
(F^{-1})_{\nu\mu} & (F^{-1})_{\nu\nu}
\end{array}
\right)^{-1}
\left(
\begin{array}{c}
\Delta x_\mu \\
\Delta x_\nu
\end{array}
\label{eqn_chi2}
\right),
\end{equation}
where $\Delta x_\mu = x_\mu - x_\mu^{\Lambda\mathrm{CDM}}$. The correlation coefficient $r_{\mu\nu}$ between two parameters is defined as
\begin{equation}
r_{\mu\nu} =
\frac{(F^{-1})_{\mu\nu}}{\sqrt{(F^{-1})_{\mu\mu}(F^{-1})_{\nu\nu}}},
\end{equation}
and describes the degree of dependence between the parameters $x_\mu$ and $x_\nu$ by assuming numerical values close to 0 for independent, and close to unity for strongly dependent parameters.

\begin{figure*}
\vspace{0.5cm}
\resizebox{0.85\hsize}{!}{\includegraphics{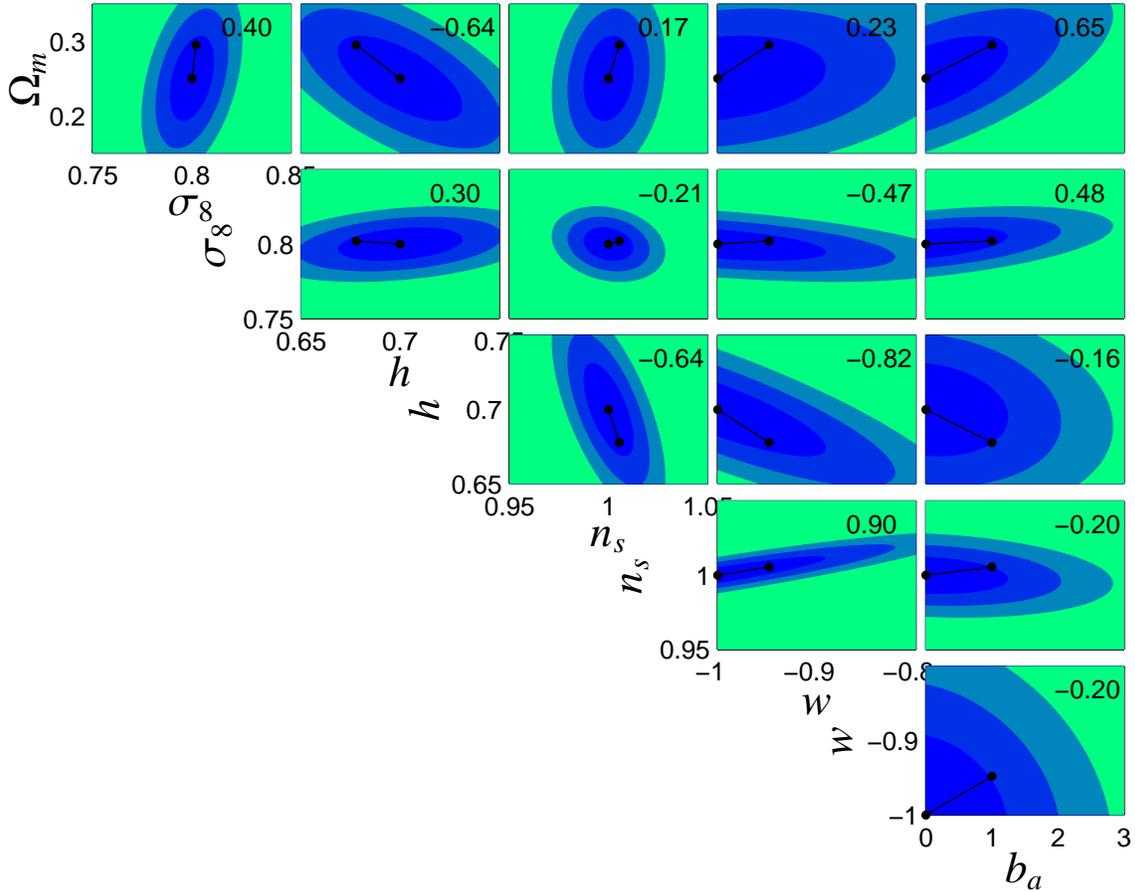}}
\vspace{0.5cm}
\caption{Constraints on the parameters $\Omega_m$, $\sigma_8$, $h$, $n_s$, $w$ and $b_a$, from a combined iSW and CMB, using $\Lambda$CDM as the fiducial model. The ellipses correspond to $1\sigma\ldots3\sigma$ intervals. Additionally, the vectors $(\delta_\mu,\delta_\nu)$ point from the fiducial cosmology to the new best-fit position and indicate the bias in the estimation of the cosmological parameters if tracer bias evolution is neglected and if one falsely assumes $b_a=0$ instead of $b_a=1$. The number in the upper right corner of each panel gives the correlation coefficient $r_{\mu\nu}$ to two digits.}
\label{fig_fisher}
\end{figure*}

The degeneracies between the cosmological parameters $\Omega_m$, $\sigma_8$, $h$, $n_s$, $w$ the bias evolution parameter $b_a$ is shown in Fig.~\ref{fig_fisher}. Most importantly, there is an anticorrelation between $b_a$ and $\sigma_8$, which both increase the iSW-signal $C_{\tau\gamma}(\ell)$, likewise for $n_s$ and $w$.

\begin{figure}
\resizebox{\hsize}{!}{\includegraphics{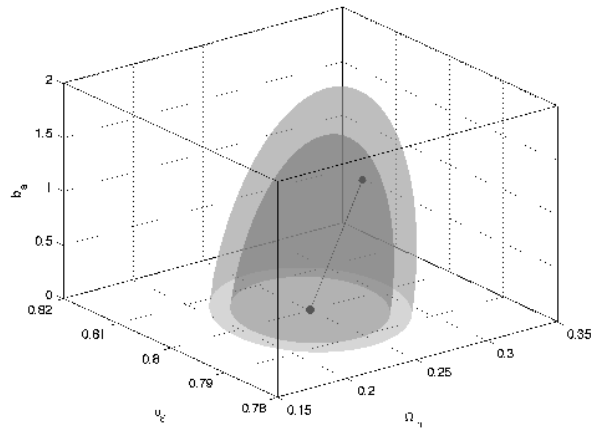}}
\caption{Fisher matrix constraint from $C_\mathrm{iSW}(\ell)$ and $C_\mathrm{CMB}(\ell)$ on the triplet $(\Omega_m,\sigma_8,b_a)$. The isoprobability surfaces correspond to $1\sigma$ and $2\sigma$. The vector $(\delta \Omega_m,\delta \sigma_8,\delta b_a)$ indicates the parameter estimation bias.}
\label{fig_degeneracy_omega}
\end{figure}

\begin{figure}
\resizebox{\hsize}{!}{\includegraphics{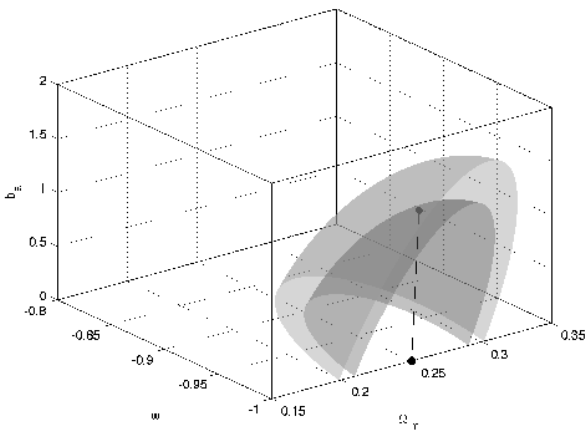}}
\caption{Fisher matrix constraint from $C_\mathrm{iSW}(\ell)$ and $C_\mathrm{CMB}(\ell)$ on the triplet $(\Omega_m,w,b_a)$. The isoprobability surfaces correspond to $1\sigma$ and $2\sigma$. The vector $(\delta \Omega_m,\delta_w,\delta b_a)$ indicates the parameter estimation bias.}
\label{fig_degeneracy_w}
\end{figure}

Figs.~\ref{fig_degeneracy_omega} and~\ref{fig_degeneracy_w} summarise the degeneracy of the cosmological parameters $\Omega_m$ and $\sigma_8$ (i.e. the strength of the gravitational potentials) and of $\Omega_m$ and $w$ (describing the particular dark energy model) on the bias evolution parameter $b_a$. It is apparent in both cases that the inclusion of $b_a$ as an additional parameter makes the parameter accuracy worse, and that the bias evolution $b_a$ can be constrained to be $<0.35$ by the iSW-data alone, with the CMB providing tight constraints on the cosmological parameters.

The errors on the individual parameters, especially the bias evolution parameter $b_a$, from a combined iSW- and CMB-measurement are given by Table~\ref{table}. These errors follow from the Cram{\'e}r-Rao bound:
\begin{equation}
\sigma_\mu = \Delta x_\mu= \sqrt{(F^{-1})_{\mu\mu}}.
\end{equation}

\begin{table}
\begin{center}
\begin{tabular}{llll}
\hline\hline
fiducial model & statistical error & estimation bias & q\\
\hline
$\Omega_m = 0.25$ 	& $\Delta\Omega_m = 0.038$	&$\delta\Omega_m = 0.045$	& q=1.19\\
$\sigma_8 = 0.8$	& $\Delta\sigma_8 = 0.007$	&$\delta\sigma_8 = -0.002$	& q=0.28\\
$h = 0.72$		& $\Delta h = 0.021$		&$\delta h = -0.022$		& q=1.07\\
$n_s = 1.0$		& $\Delta n_s = 0.008$		&$\delta n_s = 0.005$		& q=0.65\\
$w = -1.0$		& $\Delta w = 0.072$		&$\delta w = 0.052$		& q=0.72\\
$b_a = 0.0$		& $\Delta b_a = 0.824$		&$\delta b_a = 1.0$		& q=1.21\\
\hline
\end{tabular}
\end{center}
\caption{Cram{\'e}r-Rao bounds $\Delta x_\mu$ on the parameter set $x_\mu$ as well as parameter estimation biases $\delta x_\mu$ and the fraction $q=\left|\delta_\mu/\Delta_\mu\right|$ between systematical and statistical error, from a combined CMB- and iSW-measurement, with $\Lambda$CDM and constant bias as the cosmological model.}
\label{table}
\end{table}

% --- section: parameter estimation bias --- %
\section{Parameter estimation bias}\label{isw_bias}
The question of parameter estimation bias in the presence of systematics was addressed in different contexts: reionisation \citep{2005ApJ...630..657Z}, weak cosmic shear \citep{2006MNRAS.366..101H, 2007arXiv0710.5171A} and Sunyaev-Zeldovich contaminations to the CMB spectrum \citep{2009MNRAS.392.1153T}. In this section, we adapt the formalism worked out by \citep{2009MNRAS.392.1153T}, because it can be used in the regime of strong systematic contributions to the observable, in order to quantify how the interpretation of the data in cosmological model without bias evolution impacts on the estimation of cosmological parameters, when in reality the bias is evolving. We restrict the parameter space to the cosmological parameters $\Omega_m$, $\sigma_8$, $h$, $n_s$ and $w$, while assuming the bias to evolve according to $b(a)=2-a$ (i.e. $b_0=1$ and $b_a=1$), and while interpreting the data with a constant bias model $b(a)\equiv b_0$, $b_a=0$.

With our choice of the particular bias evolution model $b(a)=b_0+(1-a)b_a$ we can decompose the distance weighting functions
\begin{equation}
W_\gamma(\chi)W_\tau(\chi) = W^\prime(\chi)b(\chi), 
\end{equation}
such that the iSW cross-spectrum $C_{\tau\gamma}(\ell)=C_\mathrm{iSW}(\ell)+C_\mathrm{evo}(\ell)$ can be separated into a part with non-evolving bias $C_\mathrm{iSW}(\ell)$ and an additive systematic, containing the bias evolution $C_\mathrm{evo}(\ell)$, by identifying the two contributions
\begin{equation}
C_\mathrm{iSW}(\ell) =
b_0\int_0^{\chi_H}\dd\chi\:\frac{W^\prime(\chi)}{\chi^2}
P_{\delta\varphi}(k=\ell/\chi),
\end{equation}
\begin{equation}
C_\mathrm{evo}(\ell) =
b_a\int_0^{\chi_H}\dd\chi\:\frac{W^\prime(\chi)}{\chi^2}
\left[1-a(\chi)\right] P_{\delta\varphi}(k=\ell/\chi).
\end{equation}
Using these relations, we define the power spectrum of the true model $C_1(\ell)$ with evolving bias,
\begin{equation}
C_1(\ell) = C_\mathrm{iSW}(\ell) + C_\mathrm{evo}(\ell)
\end{equation}
as well as the spectrum of the false model $C_2(\ell)$, which neglects bias evolution,
\begin{equation}
C_2(\ell) = C_\mathrm{iSW}(\ell),
\end{equation}
where the observed spectra $\tilde{C}_i(\ell)$ are unbiased estimators of the theoretical spectra $C_i(\ell)$ in each case, because of the cross-correlation measurement method. 

If the data in reality is described by $C_1(\ell)$, but is fitted erroneously with $C_2(\ell)$, the best-fit values are biased, because the remaining parameters have to emulate the ignored degree of freedom, in our case the bias evolution $b_a$. This parameter estimation bias, defined as the distance $\bdelta\equiv\bmath{x}_2-\bmath{x}_1$ between the best-fit values $\bmath{x}_1$ of the true model $C_1(\ell)$ and $\bmath{x}_2$ of the false model $C_2(\ell)$ can be derived using the formalism by \citet{2009MNRAS.392.1153T}, who expanded the $\chi^2$-function of $C_2(\ell)$ at $\bmath{x}_1$ in a Taylor series and recovered the best-fit position $\bmath{x}_2$ by extremisation of the ensemble-averaged $\bra\chi^2\ket$:
\begin{equation}
\left\bra\frac{\partial}{\partial x_\mu}\chi^2\right\ket_{\bmath{x}_1} = 
-\sum_\nu\left\bra\frac{\partial^2}{\partial x_\mu\partial x_\nu}\chi^2\right\ket_{\bmath{x}_1}\delta_\nu.
\end{equation}
The resulting linear system of equations $\sum_\nu G_{\mu\nu}\delta_\nu = a_\mu$ can be inverted for the estimation bias $\bdelta$, $\delta_\mu = \sum_\nu (G^{-1})_{\mu\nu}a_\nu$. The two quantities $G_{\mu\nu}$ and $a_\mu$ follow from the derivatives of the $\chi^2$-function of model $C_2(\ell)$, evaluated at $\bmath{x}_1$,
\begin{eqnarray}
G_{\mu\nu}^\mathrm{iSW} & \equiv & 
\sum_{\ell=\ell_\mathrm{min}}^{\ell_\mathrm{max}}\mathrm{Cov}^{-1}\left[\frac{\partial C_\mathrm{iSW}(\ell)}{\partial x_\mu}\frac{\partial C_\mathrm{iSW}(\ell)}{\partial x_\nu} - C_\mathrm{evo}(\ell)\frac{\partial^2 C_\mathrm{iSW}(\ell)}{\partial x_\mu\partial x_\nu}\right],\nonumber\\
a_{\mu} & \equiv &
\sum_{\ell=\ell_\mathrm{min}}^{\ell_\mathrm{max}} \mathrm{Cov}^{-1}\left[C_\mathrm{evo}(\ell)\frac{\partial C_\mathrm{iSW}(\ell)}{\partial x_\mu}\right].
\end{eqnarray}
As in the case of the Fisher-analysis, the iSW-measurement is enhanced by using CMB priors on the relevant parameters. A short calculation shows that the prior information can be in corporated by the replacement
\begin{equation}
G_{\mu\nu} = G_{\mu\nu}^\mathrm{iSW} + F_{\mu\nu}^\mathrm{CMB},
\end{equation}
in the case of uncorrelated iSW- and CMB-likelihoods.

The biases in parameter estimation from the iSW-effect are depicted alongside the degeneracies in Fig.~\ref{fig_fisher}, for a bias evolution model with $b_a=1$. Table~\ref{table} lists the parameter biases from the combined CMB and iSW-measurement: The shifts of the best-fit position are most prominent for the parameters $\sigma_8$, $n_s$ and $w$, and are small for $\Omega_m$ and $h$. Generally, the estimation bias is roughly $\propto b_a$ and the inclusion of a strong prior, like in our case of $F^\mathrm{CMB}_{\mu\nu}$, reduces the estimation bias significantly. The parameter bias is comparable to the statistical uncertainty at the $1\sigma$-level for the parameters $\sigma_8$, $n_s$ and $w$, which indicates that tracer bias evolution is an important systematic. Another point worth noting is that the estimation bias is not necessarily linked to the degeneracy direction.

The estimation bias on $w$ is particularly troublesome as $w$ does not come out negative enough, and one would conclude from the measurement the presence of dark energy when in reality one would deal with a cosmological constant. $\sigma_8$ is biased towards smaller values, because bias evolution implies a higher average bias in the past, such that $\sigma_8$ can be lowered while retaining the same amplitude for the iSW-signal. Fitting the iSW-data with $b_a$ as a free parameter deteriorates the statistical accuracy especially on $\Omega_m$ and $w$, which leads to the conclusion that it should be best to include external information on $b_a$ from other data sets. Given the low significance of the iSW-signal, it is doubtful whether the data alone would prefer a more complicated model including bias evolution over a model with constant bias on grounds of Bayesian model selection \citep{2006A&G....47d..30L}.

% --- section: summary --- %
\section{Summary}\label{sect_summary}
This paper treats the influence of bias evolution on the estimation of cosmological parameters from the iSW-effect, in particular the dark energy equation of state parameter $w$. We consider a cross-correlation measurement with the galaxy density as biased large-scale structure tracer and introduce two parameters $b_0$, $b_a$ as a generic bias evolution model, and assume that at redshifts relevant for the iSW-effect, the bias is well described with $b(a)=b_0+(1-a)b_a$.
\begin{enumerate}
\item{The iSW-spectra are sensitive to evolving bias model mainly through the amplitude of the mean bias of the tracer sample, with an additional scale-dependent contribution which is most appreciable on large angular scales.}
\item{Parameter degeneracies between the cosmological parameters $\Omega_m$, $\sigma_8$, $h$, $n_s$ and $w$ on one side and the bias evolution parameter $b_a$ on the other side were investigated in a classical Fisher-matrix approach, with $\Lambda$CDM as the fiducial cosmology. Due to the weakness of the iSW-effect, we consider combined constraints from the iSW-effect and from primary CMB fluctuations. We would like to emphasise that it is not possible to separate bias evolution from parameter estimation, but on the other hand find that $b_a$ and $w$ can be constraint simultaneously from the iSW-spectrum.}
\item{In quantifying biases in the estimation of cosmological parameters from the iSW-cross spectrum $C_{\tau\gamma}(\ell)$ if the data is interpreted in a model with constant galaxy bias where in reality the bias is evolving we find that bias evolution shifts the best fit position by an amount comparable to the statistical accuracy in the case of $\sigma_8$, $n_s$ and $w$, indicating the importance of bias evolution on iSW parameter estimation. Specifically, the ratio between systematical and statistical error amounts to 1.19 in $\Omega_m$, 0.28 in $\sigma_8$, and to 0.72 in $w$ for a bias evolution parameter of $b_a=1$.}
\item{The parameters most affected are $\sigma_8$, due to the higher average bias at earlier times, $n_s$ due to the change in the shape of the cross-correlation spectrum and $w$, which is not as well constrained as e.g. $\Omega_m$ from primary CMB anisotropies. Ignoring bias evolution yields less negative values for the equation of state $w$, hinting at the danger that one might favour dark energy models over $\Lambda$CDM.}
\end{enumerate}
In a future paper we will investigate means to control bias evolution by carrying out iSW-tomography as well as by combining the iSW-signal with biased tracers such as galaxies with the weak lensing signal from the same galaxy sample as an unbiased tracer of the potential fluctuations.

% --- section: Acknowledgements --- %
\section*{Acknowledgements}
We would like to thank Carlos Hern{\'a}ndez-Monteagudo and Patricio Vielva-Mart{\'i}nez for valuable comments, Tommaso Giannantonio for help on practical issues concerning iSW-cross correlations, and Pier-Stefano Corasaniti for an exchange of ideas during the workshop on dark energy at IAS/Orsay. We thank in particular Nicolas Taburet for providing his parameter bias formalism prior to publication, and Angelos Kalovidouris for proof-reading the paper.

% --- section: SZ definitions --- %
\bibliography{bibtex/aamnem,bibtex/references}

\begin{thebibliography}{}

\bibitem[\protect\citeauthoryear{{Abramowitz} \& {Stegun}}{{Abramowitz} \&
  {Stegun}}{1972}]{1972hmf..book.....A}
{Abramowitz} M.,  {Stegun} I.~A.,  1972, {Handbook of Mathematical Functions}.
Handbook of Mathematical Functions, New York: Dover, 1972

\bibitem[\protect\citeauthoryear{{Amara} \& {Refregier}}{{Amara} \&
  {Refregier}}{2007}]{2007arXiv0710.5171A}
{Amara} A.,  {Refregier} A.,  2007, ArXiv 0710.5171, 710

\bibitem[\protect\citeauthoryear{{Bardeen}, {Bond}, {Kaiser} \&
  {Szalay}}{{Bardeen} et~al.}{1986}]{1986ApJ...304...15B}
{Bardeen} J.~M.,  {Bond} J.~R.,  {Kaiser} N.,    {Szalay} A.~S.,  1986, \apj,
  304, 15

\bibitem[\protect\citeauthoryear{{Basilakos} \& {Plionis}}{{Basilakos} \&
  {Plionis}}{2001}]{2001ApJ...550..522B}
{Basilakos} S.,  {Plionis} M.,  2001, \apj, 550, 522

\bibitem[\protect\citeauthoryear{{Blanton}, {Cen}, {Ostriker}, {Strauss} \&
  {Tegmark}}{{Blanton} et~al.}{2000}]{blanton00}
{Blanton} M.,  {Cen} R.,  {Ostriker} J.~P.,  {Strauss} M.~A.,    {Tegmark} M.,
  2000, \apj, 531, 1

\bibitem[\protect\citeauthoryear{{Boughn} \& {Crittenden}}{{Boughn} \&
  {Crittenden}}{2004}]{2004Natur.427...45B}
{Boughn} S.,  {Crittenden} R.,  2004, \nat, 427, 45

\bibitem[\protect\citeauthoryear{{Cabr{\'e}}, {Gazta{\~n}aga}, {Manera},
  {Fosalba} \& {Castander}}{{Cabr{\'e}} et~al.}{2006}]{2006MNRAS.372L..23C}
{Cabr{\'e}} A.,  {Gazta{\~n}aga} E.,  {Manera} M.,  {Fosalba} P.,
  {Castander} F.,  2006, \mnras, 372, L23

\bibitem[\protect\citeauthoryear{{Chevallier} \& {Polarski}}{{Chevallier} \&
  {Polarski}}{2001}]{2001IJMPD..10..213C}
{Chevallier} M.,  {Polarski} D.,  2001, International Journal of Modern Physics
  D, 10, 213

\bibitem[\protect\citeauthoryear{{Cooray}}{{Cooray}}{2002}]{2002PhRvD..65h3518%
C}
{Cooray} A.,  2002, \prd, 65, 083518

\bibitem[\protect\citeauthoryear{{Crittenden} \& {Turok}}{{Crittenden} \&
  {Turok}}{1996}]{1996PhRvL..76..575C}
{Crittenden} R.~G.,  {Turok} N.,  1996, Physical Review Letters, 76, 575

\bibitem[\protect\citeauthoryear{{Douspis}, {Castro}, {Caprini} \&
  {Aghanim}}{{Douspis} et~al.}{2008}]{2008arXiv0802.0983D}
{Douspis} M.,  {Castro} P.~G.,  {Caprini} C.,    {Aghanim} N.,  2008, ArXiv
  0802.0983, 802

\bibitem[\protect\citeauthoryear{{Fosalba}, {Gazta{\~n}aga} \&
  {Castander}}{{Fosalba} et~al.}{2003}]{2003ApJ...597L..89F}
{Fosalba} P.,  {Gazta{\~n}aga} E.,    {Castander} F.~J.,  2003, \apjl, 597, L89

\bibitem[\protect\citeauthoryear{{Fry}}{{Fry}}{1996}]{fry96}
{Fry} J.~N.,  1996, \apjl, 461, L65+

\bibitem[\protect\citeauthoryear{{Gazta{\~n}aga}, {Manera} \&
  {Multam{\"a}ki}}{{Gazta{\~n}aga} et~al.}{2006}]{2006MNRAS.365..171G}
{Gazta{\~n}aga} E.,  {Manera} M.,    {Multam{\"a}ki} T.,  2006, \mnras, 365,
  171

\bibitem[\protect\citeauthoryear{{Giannantonio}, {Crittenden}, {Nichol},
  {Scranton}, {Richards}, {Myers}, {Brunner}, {Gray}, {Connolly} \&
  {Schneider}}{{Giannantonio} et~al.}{2006}]{2006PhRvD..74f3520G}
{Giannantonio} T.,  {Crittenden} R.~G.,  {Nichol} R.~C.,  {Scranton} R.,
  {Richards} G.~T.,  {Myers} A.~D.,  {Brunner} R.~J.,  {Gray} A.~G.,
  {Connolly} A.~J.,    {Schneider} D.~P.,  2006, \prd, 74, 063520

\bibitem[\protect\citeauthoryear{{Giannantonio}, {Scranton}, {Crittenden},
  {Nichol}, {Boughn}, {Myers} \& {Richards}}{{Giannantonio}
  et~al.}{2008}]{2008arXiv0801.4380G}
{Giannantonio} T.,  {Scranton} R.,  {Crittenden} R.~G.,  {Nichol} R.~C.,
  {Boughn} S.~P.,  {Myers} A.~D.,    {Richards} G.~T.,  2008, ArXiv 0801.4380,
  801

\bibitem[\protect\citeauthoryear{{Ho}, {Hirata}, {Padmanabhan}, {Seljak} \&
  {Bahcall}}{{Ho} et~al.}{2008}]{2008arXiv0801.0642H}
{Ho} S.,  {Hirata} C.~M.,  {Padmanabhan} N.,  {Seljak} U.,    {Bahcall} N.,
  2008, ArXiv 0801.0642, 801

\bibitem[\protect\citeauthoryear{{Hu} \& {Sugiyama}}{{Hu} \&
  {Sugiyama}}{1994}]{1994PhRvD..50..627H}
{Hu} W.,  {Sugiyama} N.,  1994, \prd, 50, 627

\bibitem[\protect\citeauthoryear{{Huterer}, {Takada}, {Bernstein} \&
  {Jain}}{{Huterer} et~al.}{2006}]{2006MNRAS.366..101H}
{Huterer} D.,  {Takada} M.,  {Bernstein} G.,    {Jain} B.,  2006, \mnras, 366,
  101

\bibitem[\protect\citeauthoryear{{Lewis}, {Challinor} \& {Lasenby}}{{Lewis}
  et~al.}{2000}]{2000ApJ...538..473L}
{Lewis} A.,  {Challinor} A.,    {Lasenby} A.,  2000, \apj, 538, 473

\bibitem[\protect\citeauthoryear{{Liddle}, {Mukherjee} \& {Parkinson}}{{Liddle}
  et~al.}{2006}]{2006A&G....47d..30L}
{Liddle} A.,  {Mukherjee} P.,    {Parkinson} D.,  2006, Astronomy and
  Geophysics, 47, 040000

\bibitem[\protect\citeauthoryear{{Limber}}{{Limber}}{1954}]{1954ApJ...119..655%
L}
{Limber} D.~N.,  1954, \apj, 119, 655

\bibitem[\protect\citeauthoryear{{Linder} \& {Jenkins}}{{Linder} \&
  {Jenkins}}{2003}]{2003MNRAS.346..573L}
{Linder} E.~V.,  {Jenkins} A.,  2003, \mnras, 346, 573

\bibitem[\protect\citeauthoryear{{Marinoni}, {Le F{\`e}vre}, {Meneux},
  {Iovino}, {Pollo}, {Ilbert}, {Zamorani} \& {Guzzo}}{{Marinoni}
  et~al.}{2005}]{2005A&A...442..801M}
{Marinoni} C.,  {Le F{\`e}vre} O.,  {Meneux} B.,  {Iovino} A.,  {Pollo} A.,
  {Ilbert} O.,  {Zamorani} G.,    {Guzzo} L.,  2005, \aap, 442, 801

\bibitem[\protect\citeauthoryear{{McEwen}, {Vielva}, {Hobson},
  {Mart{\'{\i}}nez-Gonz{\'a}lez} \& {Lasenby}}{{McEwen}
  et~al.}{2007}]{2007MNRAS.376.1211M}
{McEwen} J.~D.,  {Vielva} P.,  {Hobson} M.~P.,  {Mart{\'{\i}}nez-Gonz{\'a}lez}
  E.,    {Lasenby} A.~N.,  2007, \mnras, 376, 1211

\bibitem[\protect\citeauthoryear{{Nolta}, {Wright}, {Page}, {Bennett},
  {Halpern}, {Hinshaw}, {Jarosik}, {Kogut}, {Limon}, {Meyer}, {Spergel},
  {Tucker} \& {Wollack}}{{Nolta} et~al.}{2004}]{2004ApJ...608...10N}
{Nolta} M.~R.,  {Wright} E.~L.,  {Page} L.,  {Bennett} C.~L.,  {Halpern} M.,
  {Hinshaw} G.,  {Jarosik} N.,  {Kogut} A.,  {Limon} M.,  {Meyer} S.~S.,
  {Spergel} D.~N.,  {Tucker} G.~S.,    {Wollack} E.,  2004, \apj, 608, 10

\bibitem[\protect\citeauthoryear{{Padmanabhan}, {Hirata}, {Seljak}, {Schlegel},
  {Brinkmann} \& {Schneider}}{{Padmanabhan} et~al.}{2005}]{2005PhRvD..72d3525P}
{Padmanabhan} N.,  {Hirata} C.~M.,  {Seljak} U.,  {Schlegel} D.~J.,
  {Brinkmann} J.,    {Schneider} D.~P.,  2005, \prd, 72, 043525

\bibitem[\protect\citeauthoryear{{Percival} \& {Sch{\"a}fer}}{{Percival} \&
  {Sch{\"a}fer}}{2008}]{2008MNRAS.385L..78P}
{Percival} W.~J.,  {Sch{\"a}fer} B.~M.,  2008, \mnras, 385, L78

\bibitem[\protect\citeauthoryear{{Pietrobon}, {Balbi} \&
  {Marinucci}}{{Pietrobon} et~al.}{2006}]{2006PhRvD..74d3524P}
{Pietrobon} D.,  {Balbi} A.,    {Marinucci} D.,  2006, \prd, 74, 043524

\bibitem[\protect\citeauthoryear{{Raccanelli}, {Bonaldi}, {Negrello},
  {Matarrese}, {Tormen} \& {de Zotti}}{{Raccanelli}
  et~al.}{2008}]{2008MNRAS.386.2161R}
{Raccanelli} A.,  {Bonaldi} A.,  {Negrello} M.,  {Matarrese} S.,  {Tormen} G.,
    {de Zotti} G.,  2008, \mnras, 386, 2161

\bibitem[\protect\citeauthoryear{{Rassat}, {Land}, {Lahav} \&
  {Abdalla}}{{Rassat} et~al.}{2007}]{2007MNRAS.377.1085R}
{Rassat} A.,  {Land} K.,  {Lahav} O.,    {Abdalla} F.~B.,  2007, \mnras, 377,
  1085

\bibitem[\protect\citeauthoryear{{Rees} \& {Sciama}}{{Rees} \&
  {Sciama}}{1968}]{rees_sciama_orig}
{Rees} M.~J.,  {Sciama} D.~W.,  1968, Nature, 217, 511

\bibitem[\protect\citeauthoryear{{Refregier} \& {the DUNE
  collaboration}}{{Refregier} \& {the DUNE
  collaboration}}{2008}]{2008arXiv0802.2522R}
{Refregier} A.,  {the DUNE collaboration} 2008, ArXiv 0802.2522, 802

\bibitem[\protect\citeauthoryear{{Sachs} \& {Wolfe}}{{Sachs} \&
  {Wolfe}}{1967}]{1967ApJ...147...73S}
{Sachs} R.~K.,  {Wolfe} A.~M.,  1967, \apj, 147, 73

\bibitem[\protect\citeauthoryear{{Sch{\"a}fer} \& {Bartelmann}}{{Sch{\"a}fer}
  \& {Bartelmann}}{2006}]{2006MNRAS.369..425S}
{Sch{\"a}fer} B.~M.,  {Bartelmann} M.,  2006, \mnras, 369, 425

\bibitem[\protect\citeauthoryear{{Smail}, {Hogg}, {Blandford}, {Cohen}, {Edge}
  \& {Djorgovski}}{{Smail} et~al.}{1995}]{1995MNRAS.277....1S}
{Smail} I.,  {Hogg} D.~W.,  {Blandford} R.,  {Cohen} J.~G.,  {Edge} A.~C.,
  {Djorgovski} S.~G.,  1995, \mnras, 277, 1

\bibitem[\protect\citeauthoryear{{Taburet}, {Aghanim}, {Douspis} \&
  {Langer}}{{Taburet} et~al.}{2009}]{2009MNRAS.392.1153T}
{Taburet} N.,  {Aghanim} N.,  {Douspis} M.,    {Langer} M.,  2009, \mnras, 392,
  1153

\bibitem[\protect\citeauthoryear{{Tegmark} \& {Peebles}}{{Tegmark} \&
  {Peebles}}{1998}]{tegmark98}
{Tegmark} M.,  {Peebles} P.~J.~E.,  1998, \apjl, 500, L79+

\bibitem[\protect\citeauthoryear{{Turner} \& {White}}{{Turner} \&
  {White}}{1997}]{1997PhRvD..56.4439T}
{Turner} M.~S.,  {White} M.,  1997, \prd, 56, 4439

\bibitem[\protect\citeauthoryear{{Vielva}, {Mart{\'{\i}}nez-Gonz{\'a}lez} \&
  {Tucci}}{{Vielva} et~al.}{2006}]{2006MNRAS.365..891V}
{Vielva} P.,  {Mart{\'{\i}}nez-Gonz{\'a}lez} E.,    {Tucci} M.,  2006, \mnras,
  365, 891

\bibitem[\protect\citeauthoryear{{Wang} \& {Steinhardt}}{{Wang} \&
  {Steinhardt}}{1998}]{1998ApJ...508..483W}
{Wang} L.,  {Steinhardt} P.~J.,  1998, \apj, 508, 483

\bibitem[\protect\citeauthoryear{{Zahn}, {Zaldarriaga}, {Hernquist} \&
  {McQuinn}}{{Zahn} et~al.}{2005}]{2005ApJ...630..657Z}
{Zahn} O.,  {Zaldarriaga} M.,  {Hernquist} L.,    {McQuinn} M.,  2005, \apj,
  630, 657

\end{thebibliography}
\bibliographystyle{mn2e}

\bsp

\label{lastpage}

\end{document}